\documentstyle[11pt,aaspp]{article}
\slugcomment{to appear in {\it The Astrophysical Journal}}

\begin{document}
\def\Ibar{{\hbox{\rlap{\hbox{\raise.25ex\hbox{-}}}I\llap{
  \hbox{\raise.25ex\hbox{-}}}}}}

\title{Millisecond Pulsars:  Detectable Sources of Continuous
Gravitational Waves?}
\author {Kimberly C. B. New, G. Chanmugam,
Warren W. Johnson  and Joel E. Tohline}
\affil {Department of Physics and Astronomy, Louisiana State University,
 Baton Rouge, LA 70803-4001}

\begin{abstract}
Laboratory searches for the detection of gravitational waves have
focused on the detection of burst signals  emitted during
a supernova explosion, but have not resulted in any confirmed detections.
An alternative approach has been to search
for continuous wave (CW) gravitational radiation from the Crab
pulsar. In this paper,
we examine the possibility of detecting CW gravitational radiation from
pulsars and
show that nearby millisecond pulsars are generally much better candidates.
We show that the minimum strain $h_c \sim 10^{-26}$ that can be detected
by tuning an antenna to the frequency of the millisecond pulsar PSR 1957+20,
with presently available
detector technology, is  orders of magnitude better than what has
been accomplished so far by observing the Crab pulsar, and within
an order of magnitude of the maximum strain that may be produced by it.
In addition, we point out that there is likely to be a population of rapidly
rotating neutron stars (not
necessarily radio pulsars) in the solar neighborhood whose spindown evolution
is driven by gravitational radiation.  We argue that the projected sensitivity
of modern resonant detectors is sufficient to detect the subset of this
population that lies within 0.1 kpc of the sun.
\end{abstract}

\keywords{ gravitation -- radiation mechanisms:  nonthermal -- stars:  neutron
-- pulsars:  general}

\section{INTRODUCTION} There are two types of signals of gravitational
radiation that are expected to be most readily detectable
from astrophysical sources:
burst signals of short duration, from sources such as a stellar merger or
a nonspherical core collapse associated with a supernova event; and
continuous wave (CW) signals from sources
such as short period, compact binary star systems or rapidly rotating,
nonaxisymmetric (or precessing) compact stars.  Most experimental
searches for gravitational radiation have focused on the detection of
burst signals.  To date, there have been no confirmed
detections of this type of gravitational radiation, although the
terminal phase of the coalescence of neutron-star binaries appears
to be a promising source for future ground-based
laser interferometers such as LIGO (Abramovici et al. 1992;
Cutler et al. 1992; Finn \& Chernoff 1993).  The detection of CW radiation from
binary star systems by a space-based interferometer has been proposed
(Faller \& Bender 1984; Evans, Iben, \& Smarr 1987; Faller et al. 1989).
Currently, though, the only experimental effort to detect CW gravitational
radiation that is underway has been pioneered by the Tokyo group which
has searched for CW emission from the Crab Pulsar (Tsubono 1991).  We
argue here that nearby millisecond pulsars are likely to be
stronger sources of CW radiation than the Crab pulsar and therefore warrant
the attention of experimental searches for gravitational radiation.

An axisymmetric object which is rotating about its minor axis will not
emit gravitational radiation because it has no time varying quadrupole
moment.  Therefore a pulsar (or any rotating neutron star, for that
matter) must be nonaxisymmetric and/or precessing
in order for it to radiate (Ferrari \& Ruffini 1969; Zimmerman 1978, 1980;
Shapiro \& Teukolsky 1983; Barone et al. 1988).  Several mechanisms for the
production of nonaxisymmetric deformations in pulsars have been suggested
(Ipser 1971), including asymmetric crystallization of the crusts
(Ruderman 1969; Ferrari \& Ruffini 1969), pressure and magnetic
stress anisotropies (Ostriker \& Gunn 1969, Ruderman 1970), and rotationally
induced instabilities (Imamura, Friedman, \& Durisen 1985).  Misalignments
between the symmetry and spin axes of pulsars might occur as the result of
electromagnetic torques due to magnetic dipole radiation, corequakes
(Pines and Shaham 1974), or encounters with neighboring stars.

Precessing, nearby millisecond pulsars have recently been put forth by
de Ara\'ujo et al. (1994) as good candidates for the detection of gravitational
radiation by upcoming interferometric detectors.  They suggest, however,
that the signals seen by these detectors due to radiation from wobbling
pulsars may be burst signals, not CW signals.  This is because the
damping time-scale for the wobble angle due to the emission of gravitational
radiation is expected to be on the order of seconds whereas the observation
time needed to observe CW sources is likely to be $\sim 10^{7}$ s.
This situation does seem likely to present itself with the first generation
of detectors because the source signals are expected to be near the limit
of detectability.
Thus, at least for some time, nonaxisymmetric deformations may be the only
channel through which pulsars produce detectable CW radiation.

The rate at which gravitational energy is radiated from a nonaxisymmetric
object
that is rotating about its minor axis with angular velocity $\omega$ is
(Ferrari \& Ruffini 1969; Shapiro \& Teukolsky 1983),
$$ \dot E_{GR} = - {32 G \over 5 c^5} {I_3}^2 \epsilon^2 \omega^6. \eqno (1)$$
This expression has been derived in the quadrupole approximation
for nearly-Newtonian sources assuming that the object has
principal moments of inertia $I_1$, $I_2$ and $I_3$,
respectively, about its three principal axes $a \gtrsim b > c$
fixed in the body frame
(Landau \& Lifshitz 1962; Misner, Thorne, \& Wheeler 1973); and
$\epsilon \equiv (a-b)/(ab)^{1/2}$ is the ellipticity in
the equatorial plane. Here, and throughout this paper, dots denote
differentiation with respect to time.
Several authors have proposed that the Crab  pulsar is
the best candidate for detecting CW radiation
(e.g., Zimmerman  \& Szedenits 1978; Tsubono 1991)
because it has the largest spin down energy flux density
$\dot E_{rot} / (4 \pi r^2)$ of all pulsars; here $r$ is the distance to the
pulsar and
$\dot E_{rot} = I_3 \omega {\dot \omega}$
($\sim 10^{38}$ erg s$^{-1}$ for the Crab pulsar)
is the rotational energy loss rate. We note, however, that for a neutron star
of
given ellipticity $\epsilon$, the loss rate due to gravitational radiation
strongly
 favors stars that are spinning more rapidly (equation [1]).
Also, if the ellipticity is caused by the rapid rotation of the
star, one might expect the value of $\epsilon$ to be higher
in stars that are rotating more rapidly.
With this in mind, we have suggested (Barker et al. 1994; see also Schutz 1995)
that
nearby millisecond pulsars (Backer et al. 1982; Fruchter, Stinebring, \&
Taylor 1988; Johnston et al. 1993)
may be stronger sources of CW radiation than the Crab pulsar (see \S 2).

One favorable aspect of CW radiation is that a resonant detector can be
tuned to the frequency of the emission.
In order to tune a cylindrical bar detector so that
its primary quadrupole mode of oscillation resonates with the
frequency of radiation from the pulsar $\omega_0 =2\omega$ (see \S 2),
the bar must have a length
$L \approx \pi c_s/\omega_0  $, where $c_s \approx 5.2 \times 10^5$ cm s$^{-1}$
is the speed of sound in a prototypical aluminum alloy bar (see \S 3).
Thus the length of the bar must be $L \sim 1.3 P_{ms}$ m,
where $P_{ms} = 10^3 (2\pi/\omega) $ is the period of the pulsar in
milliseconds.  This is prohibitively long
($L \sim 43$ m) for the Crab pulsar which has a period $P_{ms} = 33$.
Consequently, the Tokyo group (Tsubuno 1991) has used two short crossed
bars instead of a single bar.  Because better sensitivities are achievable
using a single bar, millisecond pulsars appear to be better
candidates from a detector design standpoint as well (see \S 3).

\section{THE CASE FOR MILLISECOND PULSARS}

In the quadrupole approximation, the two polarizations of the
gravitational strain amplitude $h$ that will be received by an observer
located at a distance $r$ along the rotation axis of a source are
(Misner, Thorne, \& Wheeler 1973; Kochanek et al. 1990; Rasio \& Shapiro 1992)
$$h_+ =  {G \over c^4} {1 \over r} (\ddot{\Ibar}_{xx}-
\ddot{\Ibar}_{yy}) $$
$$h_{\times} =  {G \over c^4} {2 \over r} \ddot{\Ibar}_{xy}. \eqno (2)$$
Here ${\Ibar}_{xx}, {\Ibar}_{yy}$, and ${\Ibar}_{xy}$ are cartesian components
of the reduced quadrupole moment,
$${\Ibar}_{ij} = I_{ij}-{1 \over 3}\delta_{ij} {\rm{Tr}} I, \eqno (3)$$
where $I$ is the inertia tensor and Tr$I=(I_1+I_2+I_3)$.  If the source
is a rotating, slightly nonaxisymmetric $(a \sim b, I_1 \neq I_2)$ star,
$$\ddot{\Ibar}_{xx}-\ddot{\Ibar}_{yy} = 4{\omega ^ 2}I_3\epsilon \cos
(2\omega t) $$
$$\ddot{\Ibar}_{xy} = 2{\omega ^2}I_3\epsilon \sin (2\omega t) \eqno (4)$$
(Shapiro \& Teukolsky 1983) and
$$ h_+ = 4 {G \over c^4} {\omega^2 \over r} I_3\epsilon\cos (2\omega t)$$
$$ h_{\times} = 4 {G \over c^4} {\omega^2 \over r} I_3\epsilon\sin (2\omega t).
\eqno (5)$$
Therefore, the ``characteristic'' gravitational strain amplitude
$h_c$ produced by a star that is spinning down due to the emission
of gravitational radiation is
 $$h_c \equiv 4 {G \over c^4} {\omega^2 \over r}  I_3 \epsilon  \eqno (6a)$$
$$\sim ({\dot E_{GR} \over 1.6\times 10^{38}
 erg s^{-1}})^{1/2} {1 \over \omega r},  \eqno (6b)$$
where, in the second expression, we have inserted the value of
$I_3 \epsilon$ prescribed by equation (1) and the frequency
at which the radiation will be detected is $\omega_o = 2\omega$.

\subsection{Expected Strains from Pulsars}

When a value of $I_3 =10^{45}$ g cm$^2$ is adopted (a reasonable estimate for a
$1.4 M_{\sun}$ neutron star of radius 10 km), equation (6a) can be written
in the form
$${h_c \over \epsilon} = 4.2\times 10^{-18}
\bigl(P_{ms}^2 r_{kpc}\bigr)^{-1},  \eqno(7)$$
where $r_{kpc}$ is the distance to the star in kiloparsecs.
The right-hand side of equation (7) is given entirely in terms of
observable pulsar parameters and assumes its largest value
($h_c/\epsilon \sim 10^{-18}$) for nearby millisecond pulsars like
PSR 1957+20 and PSR 0437-47 (see Figure 1).  Unfortunately, one
cannot use this expression to predict a particular strain from a
given pulsar without making some assumption about the degree to which
the pulsar possesses a nonaxisymmetric structure.

Neutron star ellipticities could conceivably be as large as the breaking strain
of their crust, which is thought to be $\lesssim 5\times 10^{-4}$ (Shapiro \&
Teukolsky 1983).  For any specific pulsar, however, an upper limit to
$\epsilon$ can be directly inferred from a measurement of
$\dot P$ if one assumes that the spindown of the pulsar is attributed solely
to the emission of gravitational radiation (Press \& Thorne 1972).  In this
situation,
$\dot E_{rot} = I_3 \omega \dot \omega$ can be equated to
$\dot E_{GR}$ as given in equation (1) and the inferred upper limit to
the ellipticity is
$$\epsilon_{GR} = 6.0 \> \bigl(P_{ms}^3 \dot P \bigr)^{1/2}, \eqno(8)$$
where, again, we have set $I_3 = 10^{45}$ g cm$^2$.  Note that, as in
equation (7), the right-hand side of this expression is given entirely
in terms of observable pulsar parameters.
Equation (8) serves only as an upper bound to the ellipticity because other
mechanisms are likely to be responsible for some portion of the observed
$\dot P$ in pulsars.  Indeed, in most normal pulsars $\dot P$
is thought to be determined primarily by energy losses due to magnetic dipole
radiation (Ostriker \& Gunn 1969; Pacini 1967) rather than by losses due
to gravitational radiation, so the true ellipticities of most pulsars
will be less (perhaps orders of magnitude less) than the value $\epsilon_{GR}$
given by expression (8).  Note that the limit on $\epsilon$
set by this expression for millisecond pulsars is very small; for example,
for PSR 0437-47 and PSR 1957+20, $\epsilon_{GR} = 3\times 10^{-8}$ and
$1.6\times
10^{-9}$, respectively (see Figure 1 and Table 1).

It proves instructive to construct a plot of the two ``observables''
$h_c/\epsilon$ vs. $\epsilon_{GR}$ for a large number of pulsars.
Figure 1 includes (as of January, 1995) all of the ($\sim 470$) pulsars
in the Caltech Pulsar Database for which
$P_{ms}, \dot P$, and $r_{kpc}$ (the distance obtained from the dispersion
measurement) were available. It should be noted,
first, that because they tend to have very small values of $\dot P$ as well as
small values of $P_{ms}$, millisecond pulsars tend to cluster in the upper
left-hand corner of this figure.  By contrast, normal
pulsars tend to cluster in the lower right-hand region of the diagram.
The lines drawn in Figure 1, with slopes of $-1$, identify the locus of
points for sources with identical values of $h_c r_{kpc}$  as measured from
Earth under the (unlikely) assumption that the spindown of all pulsars
is attributed solely to the emission of gravitational radiation, i.e., under
the
assumption $\epsilon=\epsilon_{GR}$.
For example, by this assumption,
pulsars above and to the right of the solid line would have
$h_c > 10^{-26}r_{kpc}^{-1}$ and
the Crab and Vela pulsars, in particular, would exhibit the largest
values of $h_c$ ($\sim10^{-24}$).

When confronted with a diagram like Figure 1, one is tempted to assume
that the Crab and Vela pulsars will be the brightest CW sources in the sky.
However, all that one can safely conclude is that, because these objects
exhibit large spindown rates (which are understood to be due
to magnetic dipole radiation, not gravitational radiation)
the ``observable'' $\epsilon_{GR}$ places only a very loose constraint on
the underlying structural ellipticity of these two neutron stars.
One is {\it permitted} by $\epsilon_{GR}$ to adopt ellipticity values
as large as few $\times 10^{-4}$ --- and, hence, strains approaching $10^{-24}$
---  without conflicting with measured values of $\dot P$.  However,
in light of what we know about millisecond pulsars, it seems unreasonable
to assume that these neutron stars have structural ellipticities of this
magnitude.
In particular, under the premise that nonaxisymmetric structure in neutron
stars is likely to be rotationally enhanced (see Imamura, Friedman, \& Durisen
1985),
one would not expect
$\epsilon$ to be larger in (the relatively slowly rotating) normal pulsars
than it is in millisecond pulsars. Correspondingly,
it would be difficult to imagine that the Crab and Vela pulsars
have ellipticities larger than $\epsilon \sim 10^{-8} - 10^{-9}$
(the limit set by $\epsilon_{GR}$ in millisecond pulsars) and, hence, that
they produce strains larger than $h_c \sim 10^{-29} - 10^{-30}$.

In millisecond pulsars, it is conceivable that nonaxisymmetric
distortions are induced by the observed, relatively large rotational energies
of these objects and that measurable levels of CW radiation result.
At first glance, PSR 0437-47 would appear to be the best candidate
for the detection of CW
radiation because it is the millisecond pulsar with the largest value of the
product
$(h_c/\epsilon) \epsilon_{GR}=2.6 \times 10^{-26}$.  However,
taking the conservative view that $\epsilon$ is nowhere larger
than the value of $\epsilon_{GR}$ set by PSR 1957+20 (i.e.,
$\epsilon \leq 1.6 \times 10^{-9}$), the pulsar exhibiting the largest
$h_c$ is PSR 1957+20 itself ($h_c=1.7 \times 10^{-27}$).
We consider this to be a much more reasonable estimate of likely strains
coming from the best candidates for the detection of CW radiation
than the estimates set earlier based on the observed properties of the
Crab and Vela pulsars.

\subsection{Variation of Characteristic Strain with Age} \label{Joel1}

In models that assume that the spindown of pulsars is due entirely to
magnetic dipole radiation, it has been appreciated for quite some time
that pulsar evolutionary paths can be drawn in an ``observables'' diagram
similar to our Figure 1 if one adopts a particular function prescribing
the strength of pulsar magnetic fields as a function of time (cf., Chanmugam
1992).  Similarly, an evolutionary path in the
$h_c/\epsilon$ vs. $\epsilon_{GR}$
diagram can be plotted for any individual pulsar if one assumes that the
pulsar's measured $\dot{P}$ is entirely due to gravitational radiation and
one adopts a particular function prescribing the pulsar's ellipticity
as a function of time.  To make such an evolutionary discussion relevant
to the broad class of pulsars, rather than to one pulsar at a time, we should
first remove any distance dependence from the diagram.  By analogy with
traditional discussions of stellar evolution, we define an ``absolute''
characteristic strain $H_c$ for any pulsar to be the characteristic strain
it would exhibit if it were located a distance $r_{kpc} = 1$ from the sun.
Figure 2 re-displays all the pulsar data from Figure 1 on this ``absolute''
characteristic strain scale.

First, note that if $P_{ms} = 0.5$ is associated with the maximum rotation
rate (breakup velocity) of any neutron star, then the dotted horizontal line
drawn at
$\log_{10}(H_c/\epsilon) = -16.78$ in Figure 2 represents the largest
achievable absolute characteristic strain of any neutron star, even at birth.
(The true demarcation line may, in fact, be somewhat higher or somewhat
lower than this; its correct location cannot be established until our
understanding of the equation of state of neutron star matter improves.)
The simplest function to choose for $\epsilon_{GR} (t)$ is $\epsilon_{GR}$
equals
a constant.  Then all evolutionary trajectories are vertical in Figure 2 and,
because $\dot{P}$ is positive, all trajectories are directed downward.
The rate at which a neutron star evolves along its vertical path in Figure 2
is prescribed by equation (8).  Specifically, we can write
$$P^3dP = \left ({\epsilon_{GR} \over A} \right) ^2 dt, \eqno (9)$$
where $A = 1.9\times10^5$ s$^{-3/2}$.  Then, assuming for simplicity that the
pulsar
period initially is much less than its period at time $t$, the function $P(t)$
becomes
$$P^2 = \left({2\epsilon_{GR} \over A}\right) t^{1/2}. \eqno (10)$$
Combining this with equation (7), the rate of evolution along a (vertical)
trajectory in Figure 2 is prescribed by the following expression:
$${H_c \over \epsilon} =
\left( 4.0\times 10^{-19}s^{1/2}\right) \epsilon_{GR}^{-1} t^{-1/2}. \eqno
(11)$$

According to equation (11), a population of neutron stars that have the same
age but varying ellipticity will trace out a straight line that has a slope
of -1 in Figure 2.  More specifically, the solid and dashed line in Figure 2
locate isochrones of $t = 10^8$ and $10^{10}$ years, respectively.
It is interesting to note that none of the individual pulsars plotted in
Figure 2 lie below the $10^{10}$ year, or ``Hubble line,'' isochrone.
Somewhat surprisingly, isochrones in Figure 2 exhibit the same slope (-1)
as lines of constant absolute characteristic strain (see Figure 1).
Hence, a population of neutron stars of the same age will exhibit the same
absolute characteristic strain across the entire population.
For reference, the solid and dashed lines in Figure 2 correspond to strains of
$H_c = 7.1 \times 10^{-27}$ and $7.1\times 10^{-28}$, respectively.

\subsection{A Population of Loud CW Sources} \label{Joel2}

According to models that invoke magnetic dipole radiation
to explain the spindown of pulsars, a pulsar's magnetic field strength
is related to $P$ and $\dot P$ by the expression
$B =  3.2\times 10^{19}$ G s$^{-1/2}
\left( P \dot P\right)^{1/2}$ (Chanmugam 1992).
By this relation, the extremely small value of $\dot P$ that has
been measured for the millisecond pulsar PSR $1957+20$
can be understood only if this pulsar possesses a magnetic field
that is no stronger than $1.7 \times 10^8$ G.
It seems unlikely that this field strength is unique among the
population of rapidly rotating neutron stars (not all of which are
radio pulsars) that resides in the solar neighborhood.
Among this sub-population, gravitational radiation will dominate
over magnetic dipole radiation as the principal energy loss mechanism
throughout each star's life ($\lesssim 10^{10}$ years) as long as
the stars have equatorial ellipticities $\epsilon_{GR}
\gtrsim 6 \times 10^{-9}$.  (More generally, the criterion works out
to be $\epsilon_{GR} \gtrsim 0.20 B_{12}^2$, where $B_{12}$ is the
field strength in units of $10^{12}$ Gauss.)

Given that the tightest constraint that currently has been placed on
the ellipticity of neutron stars is $\epsilon \lesssim 5\times 10^{-4}$,
set by the breaking strain of the crust, it does not seem unreasonable to
suggest
that most neutron stars with $B \lesssim 10^8$ Gauss also have structural
ellipticities $\epsilon_{GR} \gtrsim 10^{-8}$.
(As a point of reference, according to Cook (1973) the
the gravitational field of the Earth exhibits an effective ellipticity
$\sim 10^{-6}$.)
We submit, therefore, that in the solar neighborhood there exists a
family of rapidly rotating neutron stars whose spindown is driven
by gravitational radiation.
According to the preceeding discussion, even after $10^{10}$ years of
evolution we are guaranteed that this family of objects will exhibit
an absolute strain $\gtrsim 7\times 10^{-28}.$
It would seem, then, that CW detectors designed to reach this strain level are
virtually certain to detect this population of stars.

\clearpage
\section{EXPERIMENTAL FEASIBILITY}

Experiments having the sensitivity necessary to directly detect
gravitational radiation are difficult to design and conduct.
We consider here a necessary condition for the feasibility of any new
experiments designed to improve significantly the sensitivity for
detection of CW sources.  The specification of {\it sufficient}
conditions is a much broader issue requiring discussion beyond the
scope of this paper.

It is well-understood that there are two fundamental, and predictable,
noise sources that will necessarily limit the sensitivity of any
gravitational wave detector:   thermo-mechanical force
noise and sensor noise.  The first of these arises from random forces
applied to the antenna due to its coupling to a finite-temperature heat
bath; the second results from random fluctuations at the output of
the electro-mechanical devices that are used to sense any gravitationally
induced, time-varying distortions of the antenna.  For detection of burst
sources, both types of noise are important (Solomonson et al. 1993).

The situation is different if one conducts a targeted search for
CW emission from a known pulsar.  Then information from radio observations
can be used to great advantage.
For example, from the measured pulsar rotation frequency $\omega$, one knows to
what frequency $\omega_{o}=2\omega$ the antenna should be tuned; and because
$\dot{P}$ is
extremely small, one expects to receive a steady flux of gravitational
radiation at the prescribed frequency for a very long time
($P/\dot{P} \sim 10^9 - 10^{10}$ yr for millisecond pulsars).
The continuous interaction of a resonantly tuned antenna with a steady,
periodic gravitational force allows the mechanical amplitude of the
antenna to steadily grow with time.  Coherent averaging
of the motion, accounting for the observer's and source's doppler
shifts, can be used to narrow the bandwidth of the measurement to
such an extent that the sensor noise, under reasonable conditions,
can become completely negligible.   Thus it should be possible to avoid
sensor noise altogether within a narrow range of frequencies and thereby
greatly
reduce the minimum detectable strain amplitude $h_c$ compared to the
amplitude that is detectable from burst sources.   Such techniques
have been employed in the search for CW radiation from the Crab pulsar
(Tsubono 1991).

In practice, then, the mechanical motion of the antenna will have two parts:
the motion induced by the gravitational wave signal and the motion
induced by the thermo-mechanical force noise.   This remaining noise source
is simply the random Langevin force $F$, or the mechanical equivalent of
Johnson-Nyquist noise in electrical circuits.
The relative amplitude of these two contributions to the signal
can be calculated by starting from the equations of
ordinary elastic theory and including the CW gravitational
and Langevin forces as driving terms (Merkowitz \& Johnson 1994).
The elastic equation can be solved via an eigenfunction expansion
which produces one harmonic oscillator equation for each eigenmode
of the system.
In this harmonic oscillator formulation, the CW gravitational force
takes the form
$$f_{GW} = {1 \over 2}\mu L_{e} \ddot{h}_+(t)
= - {1 \over 2}\mu  L_e h_c \omega_o^2 \cos{\omega_o t}, \eqno (12)$$
where $\mu$ is the effective mass of the detector and $L_e$ is the
effective length of the mode for a particular component of the
gravitational strain tensor.
Representing the effects of $F$ by a random gravitational field $\ddot{h}_R$,
the Langevin force $f_{F}(t)$ takes the form
$$f_{F}(t) = {1 \over 2}\mu L_e \ddot{h}_R(t). \eqno (13)$$
Therefore the total force acting on the antenna can be written as:
$$f_{tot} = {1 \over 2}\mu L_e
(\ddot{h}_R (t) - h_c \omega_{o}^2 \cos{\omega_o t}). \eqno (14)$$

The spectral density $S_R$ of any random force $f_R$ is given by
$$S_R = <\tilde{f}_R^{*} \tilde{f}_R>, \eqno (15)$$
where $\tilde{f}_R$ is the Fourier transform of $f_R$ and an asterisk denotes
the complex conjugate. Setting $f_R = f_F$, and realizing that
$\tilde{\ddot{h}}_R \cong -\omega_{o}^2 \tilde{h}_R$, we deduce
$$S_F = \bigl({1 \over 2}\mu L_e \omega_o^2 \bigr)^2 S_h, \eqno (16)$$
where
$$S_h \equiv <\tilde{h}_R^{*} \tilde{h}_R>. \eqno (17)$$
But the single-sided spectral density of the Langevin force
near a resonant mode at frequency
$\omega_o$ is fixed by simple thermodynamics and the fluctuation-dissipation
theorem to be
$$S_F = 4kT\bigl({{\mu \omega_o} \over Q} \bigr), \eqno (18)$$
where $k$ is the Boltzmann constant, $T$ is the physical temperature of the
heat bath, and $Q$ is called the
mode quality factor ($Q^{-1}$ is the dissipation coefficient).
Hence, combining equations (16) and (18) we derive
$$S_h = {{16 k T} \over {L_e^2 \mu \omega_o^3 Q}} . \eqno (19)$$
This expression gives the spectral density of the noise produced
in the antenna by the Langevin force, normalized to the amplitude
of the force on the antenna due to the gravitational wave itself.

The square root of $S_h$ is referred to as the ``strain noise'' of
the antenna and can be used to define the background noise
level below which a signal of astrophysical origin will remain
undetectable.  More specifically, if one integrates the signal over
an observing time $\tau$ and demands a $4 \sigma$ confidence level
for detection, the smallest reliably detectable $h_c$ will be
$$h_c\cong 4\sqrt {{S_h \over \tau }}. \eqno (20)$$

When developing a detector for CW gravitational radiation, expression
(19) is important because it identifies design parameters that contribute
most significantly to the amplitude of the random noise near the resonant
mode of the antenna.  It is important to realize, in addition,
that the factors of length, mass,
and frequency are not independent. They are coupled by the geometry and
composition of the antenna.  For example, for the primary quadrupole mode of a
``long'' right circular cylinder, or bar, with length $L$ and diameter
$d << L$, the mode parameters become
$$\mu = {1 \over 8} \rho \pi d^2 L, \eqno (21a)$$
$$L_e \approx {4 \over \pi^2} L \approx {{4 c_s} \over {\pi \omega_o}}, \eqno
(21b)$$
where $\rho$ is the mass density of the antenna material and $c_s$ is
the speed of sound in the material.
Substituting these expressions into equation (19), the expression for
strain noise becomes
$$\sqrt{S_h} = \left( {8kT \over {d^2 \rho c_s^3 Q}}\right )^{1/2}
= 1.2 \times 10^{-23}
\left( {{{1m} \over d}}\right)
\left( {{T\over {0.05K}}} \right)^{1/2}
\left( {{{10^8} \over Q}} \right)^{1/2} {\rm Hz}^{-1/2}, \eqno (22)$$
where we have used the density ($\rho \approx 2.7$ g cm$^{-3}$)
and sound velocity of aluminum alloy ($c_s \approx 5.2 \times 10^5$ cm
s$^{-1}$),
which we regard as the most likely material for such an antenna.
Note that neither $\omega_o$ nor $L$ has
a direct effect on the sensitivity of the bar antenna, but the source frequency
does affect the length $L$ of the bar ($L=2.1$ m for PSR 1957+20, for example).

With current technology, the chosen scale parameters are within
the realm of feasibility.  The
specified diameter, for example, is at the limit of current commercial casting
capability, and $T \sim 0.05K$ has recently been achieved
for the NAUTILUS bar detector under development by the Rome group
(Astone 1991).  We have assumed that a moderate improvement in the value
of $Q$ over presently available bar antennas can be accomplished.
Such values of $Q$ have been achieved by annealing
small samples of commercially obtained aluminum alloy 5056 (Marsden 1984).
The most significant technical difficulty in achieving a strain noise as
low as $10^{-23}$ Hz$^{-1/2}$, as indicated by our parameterization of
expression (22), is likely to be vibration isolation in such an
ultra-low-temperature cryostat.

An integration time $\sim$ 4 months
seems feasible for tracking a pulsar that is well timed from radio
observations, and makes it reasonable to repeat the
measurement a number of times for confirmation.
Combining expressions (20) and (22), and setting $\tau = 10^7$ s,
the smallest detectable $h_c$ becomes $1.5 \times 10^{-26}$.
By comparison, the Tokyo group has obtained an upper limit of
$h_c \sim 2 \times 10^{-21}$ for the Crab pulsar  (Owa et al. 1988).
Current experiments they are conducting are likely to lead to
improvements of a factor 100 or so (K. Tsubono, private communication).

 A recently proposed, alternative to the cylindrical bar detector
is a ``spherical'' detector having an antenna configured as a truncated
icosahedron,
that is, an antenna with the same geometric shape as the $C_{60}$
Buckyball molecule (Johnson \& Merkowitz 1993; Merkowitz \& Johnson 1994).
An antenna of this type with a diameter $D$ exhibits 5 resonant quadrupole
modes,
each having  mode parameters of the form:
$$\omega_0 \approx 3.24 {c_s \over D}, \eqno (23b)$$
$$\mu = \rho {\pi \over 6} D^3, \eqno (23a)$$
$$ L_e \approx 0.301 D. \eqno (23c)$$
Again, the size of the antenna is fixed by the sound velocity and the
source frequency; for example, $D=2.2$ m for PSR 1957+20.
Substituting these expressions into equation (19) gives the following
expression for the strain noise of a spherical antenna:
$$\sqrt{S_h} = \left ( {{9.92 kT} \over {D^2\rho c_s^3Q}} \right )^{1/2}
= 6.3 \times 10^{-24} \left( {1.6  \over P_{ms}}\right)
\left( {{T\over {0.05K}}} \right)^{1/2}
\left( {10^8 \over Q} \right)^{1/2} {\rm Hz}^{-1/2}. \eqno (24)$$
The first equality appears to be little different from the
bar case but, actually, because the diameter $D$ for a sphere is
larger than the diameter $d$ for a bar, when both have the
same resonant frequency the value  of $\sqrt{S_h}$ for the sphere is
smaller.  Additional advantages of a spherical detector are that it can be
equally sensitive to a wave from any direction
and it is capable of measuring the direction and polarization of the wave
(Forward 1971; Wagoner \& Paik 1976; Merkowitz \& Johnson 1994).

In the second equality of expression (24) we have used the coupling of
the antenna's diameter
to the frequency to eliminate $D$ in favor of the pulsar period $P_{ms}$.
We have again assumed aluminum alloy as the material.
As in the case of the bar detector, a significant technical difficulty
will be maintaining high $Q$ when casting or joining aluminum to make a
sphere of diameter $2.2$ m designed to resonate with twice the rotation
frequency of PSR 1957+20.  If the integration time is again chosen to
be $\tau=10^{7} $s, the minimum detectable strain for a spherical detector
tuned to PSR 1957+20 is predicted to be $h_{c} = 8.0\times 10^{-27}$.

\section{DISCUSSION}
Nearby millisecond pulsars are good candidates for the detection
of CW gravitational radiation.  Because of their close proximity
and rapid rotation, they are capable of emitting radiation with
larger gravitational strain amplitudes than pulsars that are
farther away and/or have longer periods.

The minimum strain $h_c \sim 10^{-26}$
($1.5 \times 10^{-26}$ for a bar detector, $8.0 \times 10^{-27}$
for a spherical detector) that can be detected by designing an antenna
tuned to the rotation frequency of the
millisecond pulsar PSR 1957+20 and employing presently available
resonant detector technology is orders of magnitude better than what has
been accomplished so far by observing the Crab pulsar, and within
an order of magnitude of the maximum strain that can be produced
by PSR 1957+20 as a result of rotationally induced nonaxisymmetric
deformations.
The design and operation of a
resonant antenna that is tuned to the rotation frequency of PSR 1957+20 would,
at the very least, place physically meaningful constraints on the
nonaxisymmetric
ellipticity of millisecond pulsars.

We have argued (\S 2.3) that there almost certainly is a population
of rapidly rotating neutron stars (not necessarily radio pulsars)
within the solar neighborhood whose spindown evolution is driven
by gravitational radiation.
Throughout their entire lifetime, these stars will radiate at an
``absolute'' strain $H_c \gtrsim 7 \times 10^{-28}$.
It is significant that the projected sensitivity of modern resonant
detectors is sufficient to detect the subset of this population of
stars that resides within 0.1 kpc of the sun.

\acknowledgments
Support through NASA grant NAGW 2447 and NSF grant PHY-9311731 is
gratefully acknowledged.

 \clearpage

 \begin{table}

 \caption{Data for Representative Pulsars\tablenotemark{a} }

 \bigskip

 \begin{tabular} {l c c c c c}

   Pulsar   &  P$_{\rm ms}$  &$\dot{{\rm P}}$   &  r$_{\rm kpc}$  &
h$_{c}$/$\epsilon$  &    $\epsilon_{GR}$
    \\
 \tableline

  Crab         &  33.4          &   4.21E-13          &  2.49           &
1.52E-21
 &     7.55E-4
    \\

  Vela         &  89.3          &   1.25E-13          &  0.61           &
8.70E-22
 &     1.80E-3
    \\

  PSR 0437-47  &  5.76          &   1.2 E-19          &  0.14           &
9.12E-19
 &     2.89E-8
    \\

  PSR 1957+20  &  1.61          &   1.68E-20          &  1.53           &
1.07E-18
 &     1.59E-9
    \\

 \end{tabular}

 \tablenotetext{a}{From the Caltech Pulsar Database}
 \end{table}

\clearpage

 \begin{figure}
 \plotone{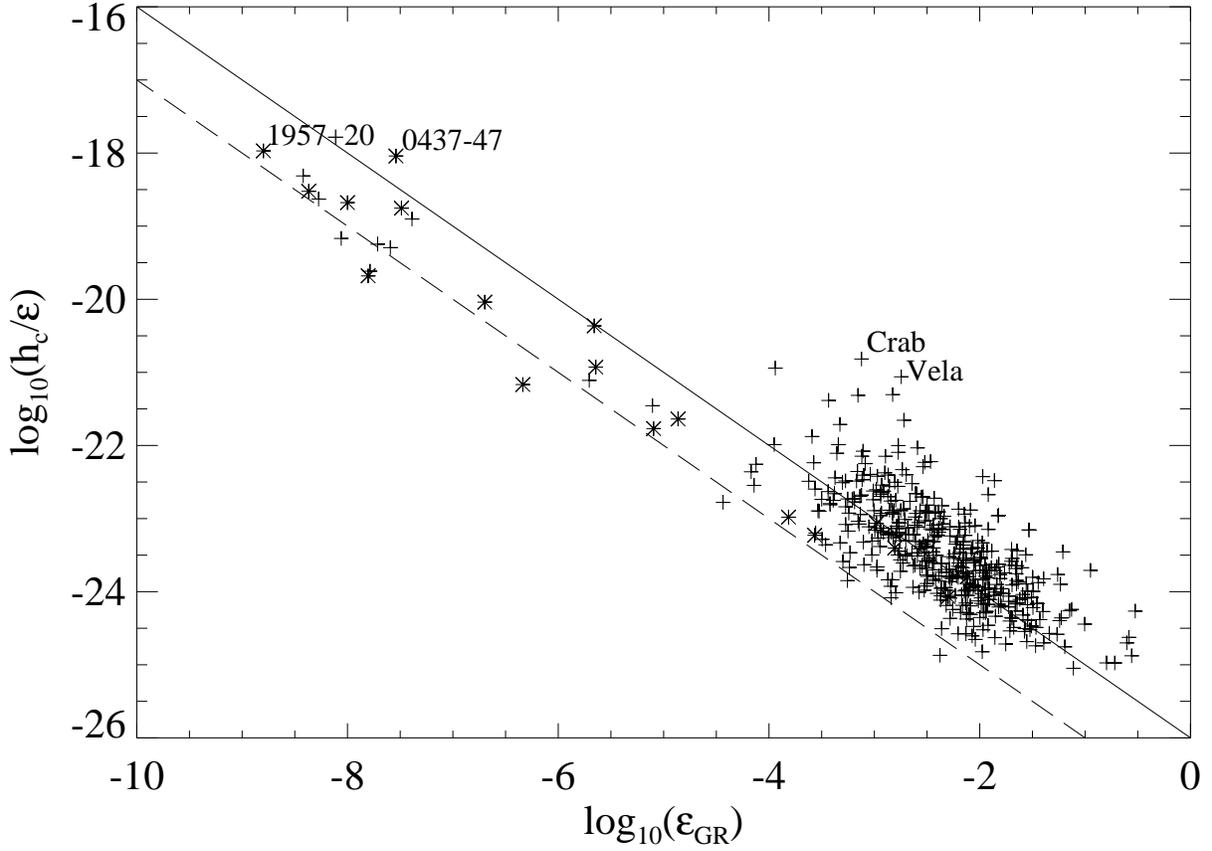}
 \caption{The two ``observables'' ${h_c/  \epsilon} = 4.2\times 10^{-18}
\bigl(P_{ms}^2 r_{kpc}\bigr)^{-1}$ versus $\epsilon_{GR} =
6.0 \> \bigl(P_{ms}^3 \dot P \bigr)^{1/2}$, for $\sim470$ pulsars from the
Caltech Pulsar Database.  Binary pulsars are denoted by asterisks.
The solid and dashed lines identify the locus of
points for sources with identical values of $h_{c}r_{kpc}=10^{-26}$ and
$10^{-27}$, respectively, under the assumption that gravitational radiation
is solely responsible for the spindown of the pulsars, and therefore, that
$\epsilon=\epsilon_{GR}$.}
 \end{figure}
 \begin{figure}
 \plotone{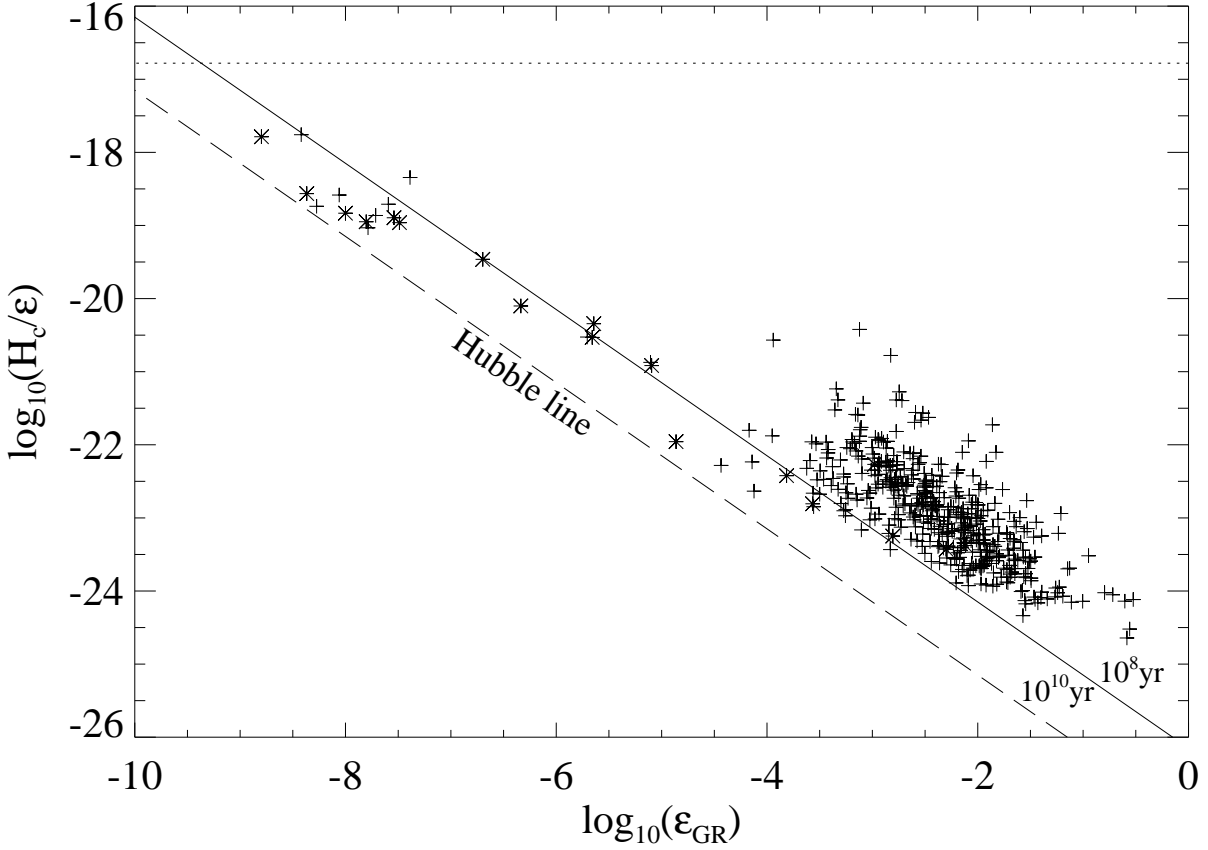}
 \caption{The same data shown in Figure 1, except that all pulsars are now
assumed to be located at a distance $r_{kpc}=1$ from the sun.  On this scale,
the ``absolute'' characteristic gravitational strain amplitude is
$H_{c} \equiv 4.2\times 10^{-18}P_{ms}^{-2}\epsilon$.  If the breakup velocity
limits
the period of a neutron star to $P_{ms}=0.5$, the dotted horizontal line marks
the
largest possible $H_{c}/\epsilon$ for any neutron star.  The solid and dashed
lines represent the isochrones of pulsars with ages of $10^{8}$ and $10^{10}$
(the Hubble line) yrs, respectively, under the assumptions that $\epsilon_{GR}$
is a constant function of time and $\epsilon=\epsilon_{GR}$.}
 \end{figure}

\end{document}